# Nonlinear Protein Degradation and the Function of Genetic Circuits

Nicolas E. Buchler[*,¶], Ulrich Gerland[#], and Terence Hwa[§]

[*]*Center for Studies in Physics and Biology, The Rockefeller University, New York, NY 10021*

[#]*Physics Department and CENS, Ludwig-Maximilians University, Munich, Germany*

[§]*Physics Department and Center for Theoretical Biological Physics, University of California at San Diego, La Jolla, CA  92093-0374*

[¶]To whom correspondence should be addressed. *e-mail: buchler@rockefeller.edu*

## ABSTRACT

The functions of most genetic circuits require sufficient degrees of cooperativity in the circuit components. While mechanisms of cooperativity have been studied most extensively in the context of transcriptional initiation control, cooperativity from other processes involved in the operation of the circuits can also play important roles. In this study, we examine a simple *kinetic* source of cooperativity stemming from the nonlinear degradation of multimeric proteins. Ample experimental evidence suggests that protein subunits can degrade less rapidly when associated in multimeric complexes, an effect we refer to as "cooperative stability". For dimeric transcription factors, this effect leads to a concentration-dependence in the degradation rate because monomers, which are predominant at low concentrations, will be more rapidly degraded. Thus cooperative stability can effectively widen the accessible range of protein levels *in vivo*. Through theoretical analysis of two exemplary genetic circuits in bacteria, we show that such an increased range is important for the robust operation of genetic circuits as well as their evolvability. Our calculations demonstrate that a few-fold difference between the degradation rate of monomers and dimers can already enhance the function of these circuits substantially.  These results suggest that cooperative stability needs to be considered explicitly and characterized quantitatively in any systematic experimental or theoretical study of gene circuits.

It is well known that the expression of genes can be regulated at a number of different stages, including transcriptional, translational, and post-translational controls. The predominant focus of many experimental and theoretical studies on genetic circuits thus far has been on the combinatorial control of transcriptional initiation, which to a large extent determines the connectivity of the circuits (1-3). Substantial efforts have also been invested in elucidating the regulatory processes that control protein degradation (4). Such processes include the covalent modification of one protein by another or the binding of one protein with another, resulting in altered turnover rates for the proteins involved (5). One consequence of an altered protein turnover rate is a shift in the steady-state cellular concentration of that protein depending on the presence/absence of other proteins. Thus, degradation control provides an additional way of establishing connections between the genes in genetic circuits. This mode of control is often used by cells when the timescale of the response is required to be short. For instance, an increase in the degradation rate allows the rapid removal of protein products (6) or conversely, protection from degradation provides a rapid way of accumulating the protein (7).

In this study, we examine an effect of protein degradation that does not involve regulatory control but can nevertheless impact the operation of genetic circuits in important ways. It is a kinetic, *cooperative* effect predicated on two essential ingredients: (i) the fact that many proteins perform their physiological functions as dimers or higher-order oligomers, and (ii) the tendency for the oligomers to be more stable (to proteolysis) than their monomeric components. This effect, referred to below as "cooperative stability", has been discussed previously in qualitative terms in the context of many well-studied examples in prokaryotes and eukaryotes; see reviews by Gottesman & Maurizi (4) and Jenal & Hennge-Aronis (5). For example, in the SOS response of *E. coli*, UmuC degradation is rescued by oligomerization with UmuD'$_2$ (8). And in *S. cerevisiae*, Johnson et al (9) showed that the dimerization of the mating-type factors *a1* and *α2* reduced the degradation rate by as much as 15-fold. Possible molecular mechanisms giving rise to cooperative stability include enhanced thermal stability of proteins upon mutual association [since thermal instability correlates with the rate of degradation (10,11)], and the burial of proteolytic recognition sequences between protein interfaces (9).

While most of the previous studies of cooperative stability focused on protein complexes with heterogeneous protein subunits, we study here its effect for typical transcription factors (TFs) that exert their biological functions only as homodimers (12). If such dimers are intended to operate at two concentrations, e.g., a "low" and a "high" concentration, then cooperative stability can widen the ratio between the two concentrations *in vivo*. This is a simple effect due to the nonlinear dependence of the protein degradation rate on the concentration, i.e., enhanced degradation rate at low concentrations due to the predominance of monomers, or alternatively, the enhanced stability of dimers that are predominant at high concentrations. Through theoretical analysis of two model gene circuits in bacteria, we will illustrate that cooperative stability can significantly enhance the function of these circuits. Specifically, we show that a several-fold effect in cooperative stability can make gene circuits more robust to stochastic fluctuations while also broadening the basin of parameter space supporting circuit function. The latter enhances the evolvability of the circuit.

**CIRCUITS AND MODELS**

In this study, we analyze two basic genetic circuits, one that displays bistability and the other spontaneous oscillation, two important classes of behavior in biomolecular circuits. The first circuit consists of only a single gene with the gene product activating its own transcription, see Fig. 1a. This is one of the simplest circuits that can produce two stable states (at LOW and HIGH levels of transcriptional activities), as was shown by theoretical analysis and an experimental implementation in *E. coli* (13). The second circuit we analyze consists of three genes connected in a ring topology, each repressing the transcription of its downstream partner, see Fig. 1b. Elowitz and Leibler (14) have shown both theoretically and experimentally that this "repressilator" can spontaneously oscillate.

    Rather than model these circuits in full detail, our goal here is to use these circuits to identify generic effects that cooperative stability imparts to their functions. Accordingly, we use simple quantitative models and describe each circuit by only a few essential parameters, so that the effects of cooperative stability can be characterized. Fig. 2 summarizes the biochemical processes considered in our model together with the associated rates. We describe the net change in the mRNA concentration due to transcription and turnover by the simple rate equation

$$\frac{dm}{dt} = \alpha \cdot g([TF]) - \lambda_m \cdot m, \qquad [1]$$

where $\alpha$ denotes the transcription rate of the promoter at full activation, $g([TF])$ describes the promoter activity as a function of the transcription factor concentration $[TF]$ (see Fig. 3), and $\lambda_m$ is the degradation rate of mRNA. Similarly, the net change in total protein concentration $p$ is

$$\frac{dp}{dt} = \nu \cdot m - (\lambda_{p1} \cdot p_1 + 2\lambda_{p2} \cdot p_2), \qquad [2]$$

where $\nu$ denotes the translation rate of mRNA. In the turnover term, the total protein concentration $p = p_1 + 2p_2$ is partitioned into monomers and dimers with concentrations $p_1$ and $p_2$, and with turnover rates $\lambda_{p1}$ and $\lambda_{p2}$ respectively. Note that the turnover term in Eq. 2 is linear in $p$ when $\lambda_{p1} = \lambda_{p2}$, and becomes nonlinear in $p$ with cooperative stability ($\lambda_{p1} > \lambda_{p2}$).

The protein products involved in our genetic circuits are all transcription factors, and as is often the case in bacteria, they function as activators or repressors only in the form of homodimers. Dimerization is assumed to be rapid, so that

$$p_2 = \frac{p_1^2}{K_d} \qquad [3]$$

with $K_d$ being the equilibrium dissociation constant.

Throughout this study, we do not explicitly include the stochastic effects of transcription, translation, and dimerization. Stochastic fluctuations are dominant when the mRNA or protein concentrations are low (19-21). Nevertheless, our formulation Eqs. 1-3 ensures (22) that the results we obtain correspond to the statistically-averaged results of more complex models that do include these stochastic effects. The advantage of our approach is that it allows us to rapidly elucidate the average behavior of each circuit for *all* combinations of its parameters. By demanding that the average protein concentrations not be too small, we can identify those (desirable) combinations of parameters for which the circuit behavior is mostly insensitive to stochastic effects.

## RESULTS

For each circuit shown in Fig. 1a and 1b, we want to identify the effect of cooperative stability on its phenotypic behavior using the quantitative model described above. Towards this end, we first determine the parameter regimes where the circuits are operational (i.e. bistable or oscillatory) without cooperative stability ($\lambda_{p1} = \lambda_{p2}$), and then with cooperative stability ($\lambda_{p1} > \lambda_{p2}$). For the latter case, we make the (reasonable) assumption that dimer turnover occurs mainly by cell growth/dilution, while monomers can experience accelerated degradation. Thus we take the typical half-life of dimers to be of the order of the cell doubling time (~ 50 min for bacteria in the exponential growth phase), whereas the typical half-life of monomers can be shortened to a few minutes.

### Bistable circuit

Since this circuit consists of a single gene with positive autoregulation, and only dimers can activate transcription, the promoter activity increases with the dimer concentration. This is indicated by the promoter activity function $g_A$ sketched in Fig. 3a, with $[TF] = p_2$. For any initial mRNA and protein concentration, this circuit will settle into a steady state given by the condition

$$\gamma \cdot g_A(p_2^*) = \lambda_{p1} \cdot \sqrt{K_d \, p_2^*} + 2\lambda_{p2} \cdot p_2^* \qquad [4]$$

where $p_2^*$ denotes the steady-state dimer concentration and $\gamma = \alpha \cdot v / \lambda_m$ is the protein synthesis rate of this gene at full activation. Eq. 4 is a simple statement that at steady state, the protein synthesis rate (left hand side) must balance the protein degradation rate (right hand side).

*Regime of bistability.* Bistability results when Eq. 4 has two stable solutions for $p_2^*$, so that the gene can settle in either a HIGH or LOW state depending on the initial condition. This property is dependent on the parameters $\gamma$, $K_d$ and the $\lambda_p$'s, as well as the *shape* of the promoter activity function $g_A(p_2)$. The activity function shown in Fig. 3a is characterized by three parameters: (i) *f*, the fold-change in promoter activity between the basal and fully-activated levels, (ii) $\kappa$, the TF concentration where the promoter activity begins to saturate, and (iii) *n*, the effective Hill coefficient which characterizes the cooperativity of the promoter activity function; see Supporting Information. Among these parameters, $\kappa$ and $\gamma$ can easily be varied over several orders of magnitude via choices of the operator (23) and the ribosomal binding sequences (24) respectively (we refer to these parameters as *programmable*). In contrast, the parameters *f*, *n*, and $K_d$ tend to be more constrained physiologically. For example, while there is no intrinsic biochemical constraint for $K_d$ to be small, values around $K_d \approx 10$ nM nevertheless appear to be typical according to *in vitro* measurements[*], e.g., $K_d \approx 20$ nM for λCI (25), $K_d \approx 8$ nM for Arc (26), $K_d \approx 10$ nM for NtrC (27), and $K_d \approx 1$ nM for Crp (28); however, see footnote in Table 1. The physiological range for all of the parameters used in our model is described in Table 1. As our goal is to study the circuit behavior for parameters within the typical physiological regime, we will analyze the circuit behavior over a wide range of values in $\kappa$ and $\gamma$, but only for a few representative values of $f$, $n$, and $K_d$.

The bistability of the circuit depends only on certain combinations of the parameters, see Supporting Information. It is revealing to plot the regime of bistability as a function of the programmable parameters, using the parameter combinations $K_d/\kappa$ and $\gamma/(\kappa \lambda_{p2})$ on the x- and y-axes, respectively, see Fig. 4a. The remaining parameters are fixed at $n = 1$ and $f = 100$, which corresponds to a strong activator (e.g., Crp) with a single operator site.

***Circuit without cooperative stability.*** For $\lambda_{p1} = \lambda_{p2}$, the corresponding bistable regime is the narrow black strip at the lower right corner of Fig.4a. With a given value of $K_d$, the black region defines an acceptable range of $\gamma$ for each value of $\kappa$. The circuit behavior *within* the bistable regime is demonstrated in Fig. 4b by plotting the steady-state monomer and dimer concentrations (grey and black curves respectively) in the HIGH and the LOW state (solid and dashed lines) as a function of $\kappa$, using the typical dimer dissociation constant of $K_d = 10$ nM. Note that the steady-state TF concentrations are very low, not exceeding $p_2^* \sim 10$ nM, which corresponds to about 10 molecules per bacterial cell. Such low concentrations are difficult to maintain reliably in the cell[†], and the circuit will be susceptible to various sources of stochastic fluctuations (40). Similar results are obtained using our model with $f = 11$ and $n \approx 1.7$, which mimics the bistable circuit studied experimentally by Isaacs et al (13); see Fig. S2 and S4 in Supporting Information. We note that strong fluctuations were indeed observed in that experiment.

---

[*] The effective value of $K_d$ *in vivo* is expected to increase with respect to its *in vitro* value due to dimer turnover. The amount of increase depends on the dimer association and dissociation rates $k_a$ and $k_d$. We find $K_d = (k_d + \lambda_{p2})/k_a$ instead of the usual *in vitro* expression $K_d = k_d/k_a$. However, for typical small proteins, this increase is estimated to be less than ~ 1 nM and hence not important for this discussion; see Supporting Information.

The steady-state protein concentrations could in principle be increased (consequently reducing stochastic effects) for larger values of $K_d$, e.g. $K_d = 1000$ nM as shown in Fig. 4c. In this case, however, the concentrations of the *nonfunctional* monomers (grey curves) are significantly larger than the concentrations of functional dimers (black curves). Indeed, without cooperative stability, monomer overproduction is a *generic* consequence of leveraging cooperativity from dimerization since the system can only exploit this source of cooperativity when the total protein concentration is much less than $K_d$ (i.e. when the protein exists primarily in monomer form). While the overproduction of monomers may or may not be detrimental to the cell for an individual gene, the monomer "load" can become a significant problem if weak dimerization is a *generic* strategy widely adopted by the cell, e.g., if *every* regulatory gene contributes 10 ~100 nonfunctional monomers in the LOW or HIGH states. This observation is a conceivable explanation for the small $K_d$ values found for typical dimeric proteins.

*Effect of cooperative stability.* The grey and the hatched areas in Fig. 4a show how the bistable parameter regime is shifted when cooperative stability is introduced with $\lambda_{p1}/\lambda_{p2} = 3$ and $\lambda_{p1}/\lambda_{p2} = 10$, respectively. For a given value of $K_d$, cooperative stability leads to a shift in the bistable regime towards an increased rate of protein synthesis $\gamma$. By increasing the protein

---

[†] Vilar & Leibler suggest that with DNA looping it may be possible to reduce this noise even with low numbers of molecules (39)

concentration, an increased $\gamma$ reduces the susceptibility of the bistable circuit to stochastic fluctuations. This increase is seen explicitly in Fig. 4d, where we plot again the steady-state monomer and dimer concentrations in the HIGH and LOW states with the typical $K_d = 10$ nM as in Fig. 4b, but this time with 10-fold cooperative stability ($\lambda_{p1} = 10 \cdot \lambda_{p2}$). Comparison of Fig. 4d to Fig. 4c demonstrates a second beneficial effect of cooperative stability: a significant reduction of the monomer load.

The origin of the beneficial effects of cooperative stability is that it extends the regime of protein concentrations where cooperativity is obtained through dimerization from $p < K_d$ where most of the proteins are monomers, to $p < K_d \cdot (\lambda_{p1}/\lambda_{p2})^2$, see Eq. S12 in Supporting Information. This results in an effective *decrease* in the value of $K_d$ needed for bistability by a factor of $(\lambda_{p1}/\lambda_{p2})^2$. The grey and hatched regions in Fig. 4a can be viewed as the black region with an appropriate reduction in $K_d$. Moreover, for the case shown in Fig. 4d with $K_d = 10$ nM and 10-fold cooperative stability, the dimer concentrations (black curves) are identical to those in Fig. 4c, which were obtained for $K_d = 1000$ nM and $\lambda_{p1} = \lambda_{p2}$. The same black curve could also be obtained for $K_d = 100$ nM and ~ 3-fold cooperative stability (not shown). Due to this large $(\lambda_{p1}/\lambda_{p2})^2$-fold reduction in $K_d$, we expect a few-fold effect in cooperative stability to exert a large impact on circuit function.

**Three-gene oscillator**

In the repressillator circuit of Fig. 1b, we have 3 genes whose mRNA and protein products, $m^{(i)}$ and $p^{(i)}$ with $i \in \{1,2,3\}$, are described by Eqs. 1-3. To focus on the generic behavior of this circuit, it is useful to assume that all 3 genes have identical properties, i.e. the same promoter structure, synthesis/turnover rates, and dissociation constant $K_d$ for protein dimerization (14). The form of the repressive promoter activity $g_R$ is shown in Fig.3b, and is characterized again by the three parameters $f$, $\kappa$, and $n$. We will examine in detail promoters with a single repressive operator site with $n = 1$; see Fig.1e.

*Oscillatory regime.* The regime of parameter space supporting oscillation is obtained by linear stability analysis around the steady-state solution with $\{dm^{(i)}/dt = 0\}$ and $\{dp^{(i)}/dt = 0\}$; see Supporting Information for details. Generally, oscillation is favored when (i) the fold-change *f* is large, (ii) the protein and mRNA turnover rates are comparable, and (iii) there is a large cooperativity/nonlinearity in synthesis and/or degradation; see also ref. (14).

*Circuit without cooperative stability.* For $\lambda_p \equiv \lambda_{p1} = \lambda_{p2}$, we show in Fig. 5a the oscillatory parameter regime for typical ($f = 100$) and exceptionally strong ($f = 1000$) repressors, paired with either typical ($\lambda_p/\lambda_m = 0.1$) or rapid ($\lambda_p/\lambda_m = 1$) protein turnover. Even for the most favorable combination, i.e. $\lambda_p/\lambda_m = 1$ and $f = 1000$, oscillation is not possible until $K_d/\kappa > 100$. Thus, for typical $K_d \sim 10$ nM and typical κ of 1~1000 nM, this system cannot sustain oscillations. In the experiment of Elowitz & Leibler (14), oscillation was obtained by (i) adding ssr-tags to the TFs so that they degrade much faster to make the protein and mRNA turnover rates comparable, and (ii) using some of the strongest repressive promoters known (30). The latter not only makes $f > 1000$, but the multiple repressive operator sites (shown in Fig. 1f) increase the value of the Hill-like coefficient associated with the transcription initiation to $n \approx 1.63$ (see Supporting Information), thereby further broadening the accessible parameter space. While the solution found by Elowitz & Leibler (14) is a triumph of synthetic engineering, we believe that such "extreme" solutions will be difficult to find by natural evolution due to the rarity of existing circuit components (e.g., promoters and TFs) with such extreme characteristics. The circuit would be much more *evolvable* if oscillation can already occur for typical components, e.g., for TFs with $\lambda_p/\lambda_m : 0.1$ and promoters consisting of a single repressive operator site with $f \leq 100$.

*Effect of cooperative stability.* By destabilizing the monomers with respect to the dimer species such that ($\lambda_{p1} > \lambda_{p2}$), cooperative stability will have two effects: First, as in the case of the self-activating one-gene switch, it extends the regime where cooperativity is obtained through dimerization from $p < K_d$ to $p < K_d \cdot \left(\lambda_{p1}/\lambda_{p2}\right)^2$, thereby making the oscillatory regime more accessible to TFs with smaller $K_d$. Second, the condition for oscillation favors the *monomer* degradation rate be closer to the mRNA degradation rate; see Supporting Information. For typical gene products with $\lambda_{p2}/\lambda_m \sim 0.1$, a 10-fold effect in cooperative stability puts $\lambda_{p1}/\lambda_m$ close to unity, thereby further extending the oscillatory regime. This is illustrated in Fig. 5b, where we specifically plotted the oscillatory regime for $\lambda_{p1} = \lambda_m = 10\lambda_{p2}$ with both $f = 100$ and $f = 1000$. In comparison to the circuit without cooperative stability (Fig. 5a), oscillation is now possible with typical molecular components.

## DISCUSSION

**Advantages of cooperative stability**

In this study, we examined the function of simple genetic circuits for dimeric TFs with various degrees of cooperative stability, i.e., where monomers turnover more rapidly than the functional dimers. In the absence of cooperative stability, the desired operation of both the one-gene switch and the 3-gene oscillator requires parameters that are on the edge of what is physiologically realizable. These limitations can be understood in simple terms: Proper functions of most biological circuits require a sufficient degree of cooperativity in the circuit components. Cooperativity at the transcription initiation stage (controlled by the fold-change *f* and the Hill coefficient *n* in our model of transcriptional control) is usually quite limited. It is thus crucial to harness cooperativity from the other processes involved in the operation of gene circuits. A simple and direct source of cooperativity that does not involve additional genes and proteins is the nonlinearity in TF dimerization. In order to harness this cooperativity, however, it is necessary to maintain the cellular TF level at or below the dimer dissociation constant $K_d$. This would leave the system with two undesirable options: Either use typical dimeric TFs with $K_d \sim 10$ nM and maintain the TFs at a very low level (e.g., below 10 molecules per cell) which leaves the system vulnerable to stochastic fluctuations, or use TFs with large $K_d$ to maintain a higher TF level (e.g., ~100 nM) to reduce these fluctuations, but expose the system to an increased load of nonfunctional monomers. We have shown that cooperative stability removes the link between the cellular TF levels and the $K_d$ values. This makes it possible to *simultaneously* maintain a cellular TF level that is robust to fluctuations and allow the circuit to harness dimer cooperativity for the typical (strong) dimers; the latter relieves the system of the monomer load problem.

Importantly, cooperative stability has broadened the parameter space for desired circuit operations, so that these circuits may be put together using typical components (TFs and promoters) without resorting to rare components with extreme properties. We suggest that this broadening of the operable parameter space is not only useful in relaxing the design constraints in synthetic biology experiments, but is moreover crucial for such circuits to emerge from natural evolution: Evolvability of a circuit requires that before selection can exert any effect, it should be possible for the organism to spontaneously assemble a primitive circuit that can sustain some rudimentary operation conferring some limited fitness advantage (41). This possibility is much enhanced if the circuit can operate using components widely accessible to the cells. Of course, cooperative stability is not the only strategy to boost the degree of cooperativity needed for circuit operations. There exist alternative strategies which may provide stronger cooperativity, including nonlinear feedback at the level of transcriptional and translational initiation/termination as well as proteolytic control involving post-translational modifications. However, such processes require additional genes and proteins. They may be the final outcome of extended refinement of genetic circuits through a prolonged evolutionary process. In contrast, cooperative stability does not require any additional molecular components except for the dimeric protein itself. Moreover, as we will argue below, cooperative stability is itself a readily evolvable molecular trait for typical dimeric proteins. Therefore, it may be used at early stages of evolution to provide a circuit with some rudimentary functions beneficial to the host, so that selection can begin to exert some effect.

It is also interesting to note that for the genetic circuits studied, cooperative stability should result in an increase in the robustness of circuit function to stochastic environmental fluctuations as well as an increase in the operational parameter space. The latter will make the circuit more robust to mutational perturbations. The correlation between robustness against environmental and genetic perturbations, known as "congruence", has been proposed on general theoretical ground (42, 43) but lacks direct experimental studies. Cooperative stability might provide a concrete molecular system to study such phenomena. Conversely, congruence implies that selection for the robustness of circuit function would naturally favor the use of proteins with cooperative stability due to the enhanced evolvability of the resulting circuit.

**Molecular mechanisms for cooperative stability**

It remains to address how broadly cooperative stability may occur in nature, or alternatively put, how readily can cooperative stability be implemented molecularly if needed. We already mentioned in the Introduction a number of specific examples where oligomerization provides protection against degradation. A recently characterized example involving homodimeric bacterial TF is TraR, a LuxR-family regulator whose homodimerization appears to require the autoinduer AI (44,45). Concomitantly, TraR is found to degrade rapidly *in vivo* in the absence of the AI, but is stabilized (with over 30x longer half-life) in the presence of the AI (44). Here we suggest that TraR is just one example of a large class of proteins that are prone to exhibit cooperative stability.

We note that many regulatory proteins are *natively unfolded*, and become folded and thermally stable only upon association with their targets (46). These include dimeric TFs which only fold upon dimerization and are referred to as "two-state dimers" (47). The best studied among this class of proteins is the Arc repressor of phage P22 (26). As have been recently characterized (48,49), conspicuous molecular features of these two-state dimers include the large number of inter-monomer contacts (compared to intra-monomer contacts) and the hydrophobicity of the interfacial contacts. We conjecture that the two-state dimers are ideal molecules for cooperative stability. First, unfolded monomers are generally believed to be more susceptible to generic degradation (10,11). Second, their exposed hydrophobic surface patches can be targeted by various proteases (33). Thus, disordered monomers may be an elegant way for nature to keep the LOW state low, while not disturbing the stable oligomers in the HIGH state.

The existence of a substantial number of two-state dimers suggests that these molecules may be readily obtained evolutionarily should they be needed. This is supported by recent experiments where upon deletion of a few residues at the end of the peptide, stably folded proteins can easily become natively unfolded (i.e., unfolded by itself), yet still fold upon target presentation (K. Plaxco, personal communication). This finding is consistent with the expectation that peptide termini away from the interaction surface contribute toward the stability of the monomers but not the complex. Together with our conjectured relation between two-state dimers and cooperative stability, it suggests that the degree of cooperative stability of dimeric proteins is readily tunable by simple terminal deletions. Thus, two-state dimers might be the favorite component of evolution to assemble gene circuits or other regulatory systems.

**Ramifications for experiment and modeling**

Cooperative stability (i.e., $\lambda_{p1} > \lambda_{p2}$) has certainly been used previously in genetic circuit modeling, both in oscillators and bistable switches. In Drosophila, reduced protein degradation by multimerization has been suggested to play a significant role in the genetic circuit controlling circadian rhythm (50,51). In the modeling of phage λ's lysis/lysogeny system, cooperative stability was implicitly used but not always in consistent ways. For example, in their modeling of entry to lysogeny, Arkin et al (52) assumed that CI monomers were degraded with half-life of ~15 min, and Cro monomers were degraded with half-life of ~ 5 min, while leaving the long-lived dimers to dilution by cell growth. With a cell doubling time of ~30-50 min, these assumptions implicitly invoke cooperative stability with $\lambda_{p1}/\lambda_{p2} \approx 2-3$ for CI and $\lambda_{p1}/\lambda_{p2} \approx 6-10$ for Cro. On the other hand, in their analysis of the experimental results of Little et al (53) [on the robustness of lysogeny to changes in the affinity of the CI-operator binding], Aurell et al (54) and Zhu et al (55) assumed that monomers and dimers were degraded with equal rates. These authors concluded that additional source(s) of cooperativity are needed to explain the observed robustness. While the recent discovery of CI octamerization at $P_{RM}$ may provide some of the necessary cooperativity (1,56,57), we suggest that cooperative stability might be another possible source of cooperativity that needs to be examined critically. In fact, Reinitz & Vaisnys (58) had speculated long ago that concentration-dependent degradation of Cro (as suggested by the data of Pakula & Sauer (59)) might provide some of the cooperativity needed to reconcile the discrepancies between theory and experiment. Given the strong impact that even a modest degree of cooperative stability can make on the phase diagram and the circuit stability (see Figs. 4 and 5), knowledge of the monomer/dimer turnover rates is crucial to guide quantitative studies of the λ-switch and to help resolve puzzles regarding the stability and robustness of lysogeny.

More generally, we note that in the current genetic circuit modeling literature, the lack of attention paid to monomer/dimer degradation has resulted in various models being unintentionally locked into forms that assume either no cooperative stability (i.e., $\lambda_{p1} = \lambda_{p2}$) or extreme cooperative stability (i.e., $\lambda_{p2} = 0$). We hope that our demonstration of the large impact of cooperative stability on circuit operation will motivate the modeling community to be more attentive to the quantitative treatment of degradation in future studies. On the experimental side, the addition of ssrA-tag on proteins has been widely used to control the turnover rate and hence the time scale of synthetic genetic circuits (14,60,61). However, depending on the degradation mode of the ssrA-tagged proteins (e.g., whether the monomers are preferentially degraded, or whether the turnover of both monomers and dimers are equally enhanced), ssrA-tagging could produce unexpected effects on the function of the circuits.

Clearly, cooperative stability needs to be investigated quantitatively in the context of specific transcriptional systems in order to establish the extent this mechanism is used in nature. Here we stress that the ubiquity of molecules (i.e., the two-state dimers) with potential for cooperative stability and the magnitude of impact this effect can exert on the operation of genetic circuits make it necessary to characterize and address its possible effects systematically in any experimental or theoretical study of genetic circuits. The generic effect of cooperative stability described here, e.g., in reducing the basal level of active proteins and consequently amplifying control signals, provides a striking example of how biophysical properties at the molecular scale can directly impact high-level function of biomolecular networks. The effect is not limited to the nonlinear degradation of dimeric transcription factors or to the specific context of genetic circuits. We expect similar nonlinear effects to extend to protein modifications in signaling networks and co-regulated components of multimeric protein complexes.

**Acknowledgements**. We are grateful to R. Bundschuh, F.R. Cross, J. Finke, S. Leibler, K. Levy, M. Louis, K. Plaxco, M. Ptashne, R. Sauer, E. Siggia, J. Vilar, J. Widom, L.C. You, and M.W. Young for helpful comments. We are especially to W.F. Loomis for encouragement and suggestions throughout the course of this work. TH acknowledges support by the NSF through grants 0211308, 0083704, 0216576, 0225630 and by a Burroughs-Wellcome functional genomics award, NEB by an NSF bioinformatics fellowship, UG by the Emmy Noether fellowship.

**Table 1**: Summary of important parameters in our model together with their typical *in vivo* values in bacteria, and the principal molecular properties that determine their values. The entries shown in bold face refer to the parameters that are programmable over a wide range. The last column lists exemplary references for the indicated physiological parameter range.

| | description | typical *in vivo* values | principal molecular determinants | references |
|---|---|---|---|---|
| $f$ | maximum fold-change in promoter activity | 10~100 (activator) 10~1000[a] (repressor) | strength of TF-RNAp interaction | (29,30) |
| $n$ | cooperativity in promoter activity | 1 ~ 2 | number of operator-bound TFs interacting with RNAp | (31) |
| $\kappa$ | **TF-operator dissociation constant** | **1~1000 nM**[b] | **operator sequence,** binding interface of TF | (23), and refs therein |
| $\lambda_m^{-1}$ | mRNA half-life | ~ 5 min | | (32) |
| $\lambda_p^{-1}$ | protein half-life | ~ 50 min (dilution) ~ few min (proteolysis) | growth rate (dilution) protein stability, degradation tag (proteolysis) | (4,33) |
| $\gamma$ | **protein synthesis rate at full activation** | **0~100 nM/min**[c] | **ribosome binding site,** transcriptional efficiency | (24,34) |
| $K_d$ | dimer dissociation constant | ~ 10 nM [d] | monomer-monomer affinity | (25-28) |

[a] Because repression involves TF-RNAp exclusion (a much stronger type of interaction than the weak attraction between RNAp and activators), the achievable fold-changes in repression can readily be much larger, e.g., $f \sim 1000$.

[b] The magnitude of $\kappa$ can be tuned by changing the number of bases matching the sequence for optimal TF-operator binding; it is an example of *programmable parameters* which play important roles in natural evolution and synthetic design of promoters (3,23).

[c] Another programmable parameter is $\gamma$ through choice of the ribosome-binding site (RBS). Maximum protein synthesis rate is limited by the rate of elongation of the ribosome. In bacteria, the elongation rate of ribosome is ~20 codons/sec and ribosomes occlude ~10 codons (34). Thus, the absolute maximum rate of protein synthesis per mRNA is ~120 proteins/min. Typical genes in bacteria have on average ~2 mRNA per cell (32), so that even with an optimal RBS, the maximum $\gamma$ is less than ~240 nM/min. For proteins that are diluted through cell division (~50 mins), this range of $\gamma$ produces steady-state protein concentrations of 0~10,000 nM.

[d] A number of bacterial TFs, e.g., BlaI (35), FIS (27), and CopR (36) have $K_d$ in the μM range. However, they are all "atypical" regulatory proteins whose *in vivo* concentrations in the active state exceed the order of 10 μM (35-37). However, it is generally believed that the over-expression of many TFs can be deleterious to the cells. For example, the maximum concentration for typical TFs in bacteria is not much more than ~100 nM; even the global regulator Crp is present only at ~1,500 nM (38). Thus, typical bacterial TFs tend to be at lower concentrations *in vivo* and tend to have smaller $K_d$.

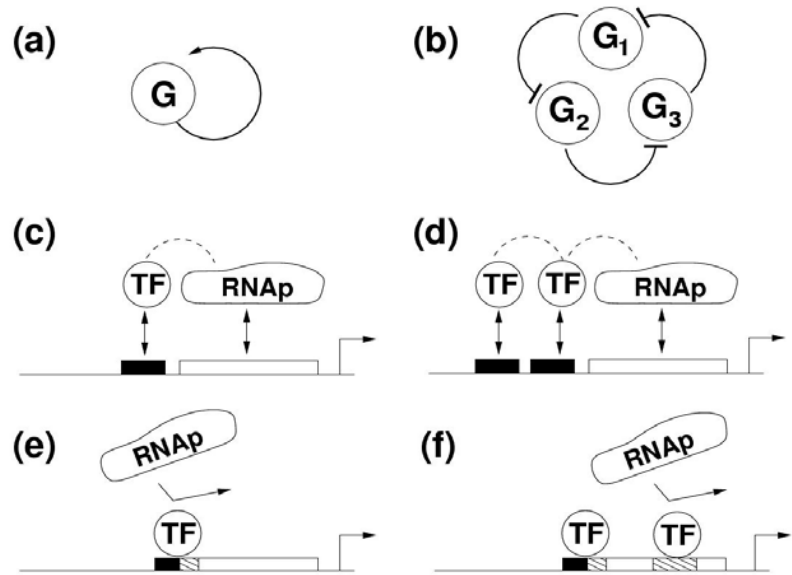

**Figure 1:** Two simple genetic circuits capable of (a) bistability and (b) oscillation. Genetic circuits consist of genes (drawn as circles) that regulate the transcriptional activity of one another. This regulation can be activating (arrow) or repressive (blunt line). Exemplary cis-regulatory architectures in bacteria using (c) one or (d) two operator sites for activation, and using (e) one or (f) two operator sites for repression. The core promoter to which RNA polymerase (RNAp) binds and the operator sites to which the transcription factors (TF) bind are drawn as open or black boxes, respectively. The dashed lines depict cooperative interaction between regulatory proteins, whereas overlapping operators (indicated by dashed boxes) denote repression mediated through excluded volume interaction.

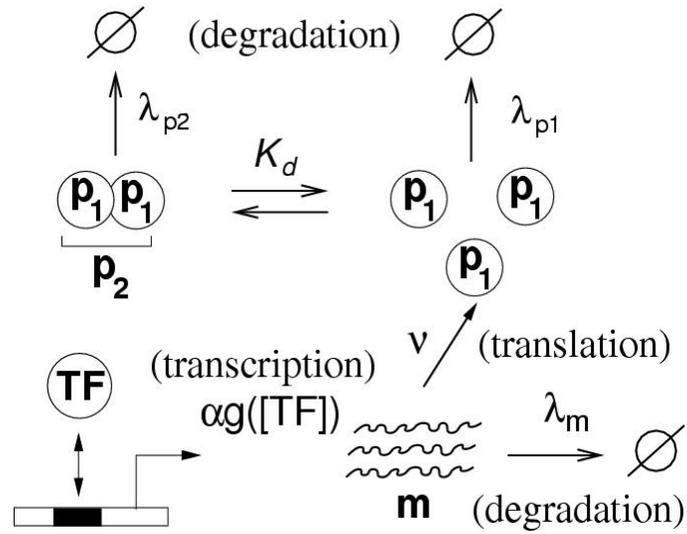

**Figure 2:** Schematic diagram of the basic parameters involved in transcription, translation, degradation, and dimerization. Transcription is governed by the transcription rate $\alpha \cdot g([TF])$ and $\alpha$ is the mRNA synthesis rate at full activation. Each mRNA is translated into protein monomer at a rate $\nu$, and degraded at a rate $\lambda_m$. The cellular concentrations of monomers ($p_1$) and dimers ($p_2$) are related by the dimer dissociation constant $K_d$. The protein degradation rate can be different for monomers ($\lambda_{p1}$) and dimers ($\lambda_{p2}$).

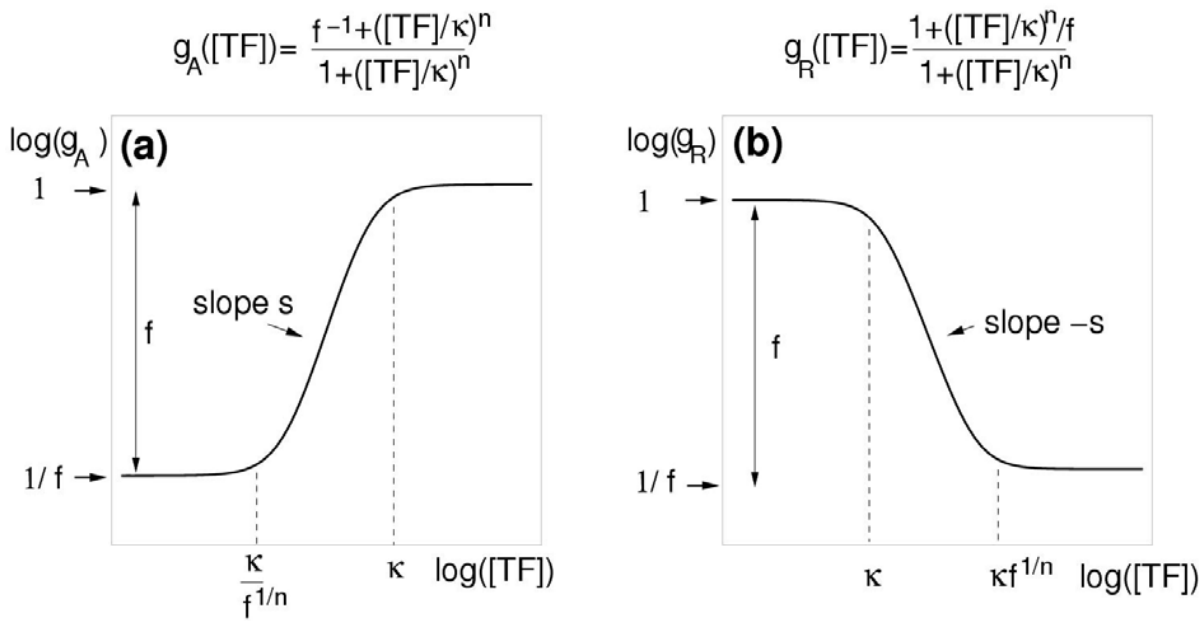

**Figure 3:** Log-log plot of the relative promoter activity $g([TF])$ versus the transcription factor concentration $[TF]$, for (a) activation and (b) repression. The general expression for the promoter activity function is written above each plot. The peak activity of a promoter is defined to be 1, the fold-change between LOW and HIGH plateaus is described by $f$, and the DNA-binding dissociation constant of a TF for its operator $\kappa$ is the concentration which separates the HIGH plateau from the transition region. The log-log slope ($s$) of the transition region (referred to as "sensitivity" in the signal transduction literature (15)) quantifies the degree of cooperativity in transcriptional control. It is controlled by the Hill coefficient $n$ and the maximum fold-change $f$, with maximum $s$ approaching $n$ for large $f$; see Supporting Information. Both (a) and (b) are approximations to promoter activity functions derived from detailed thermodynamic treatment of transcription initiation; see Supporting Information and refs. (3,16-19).

.

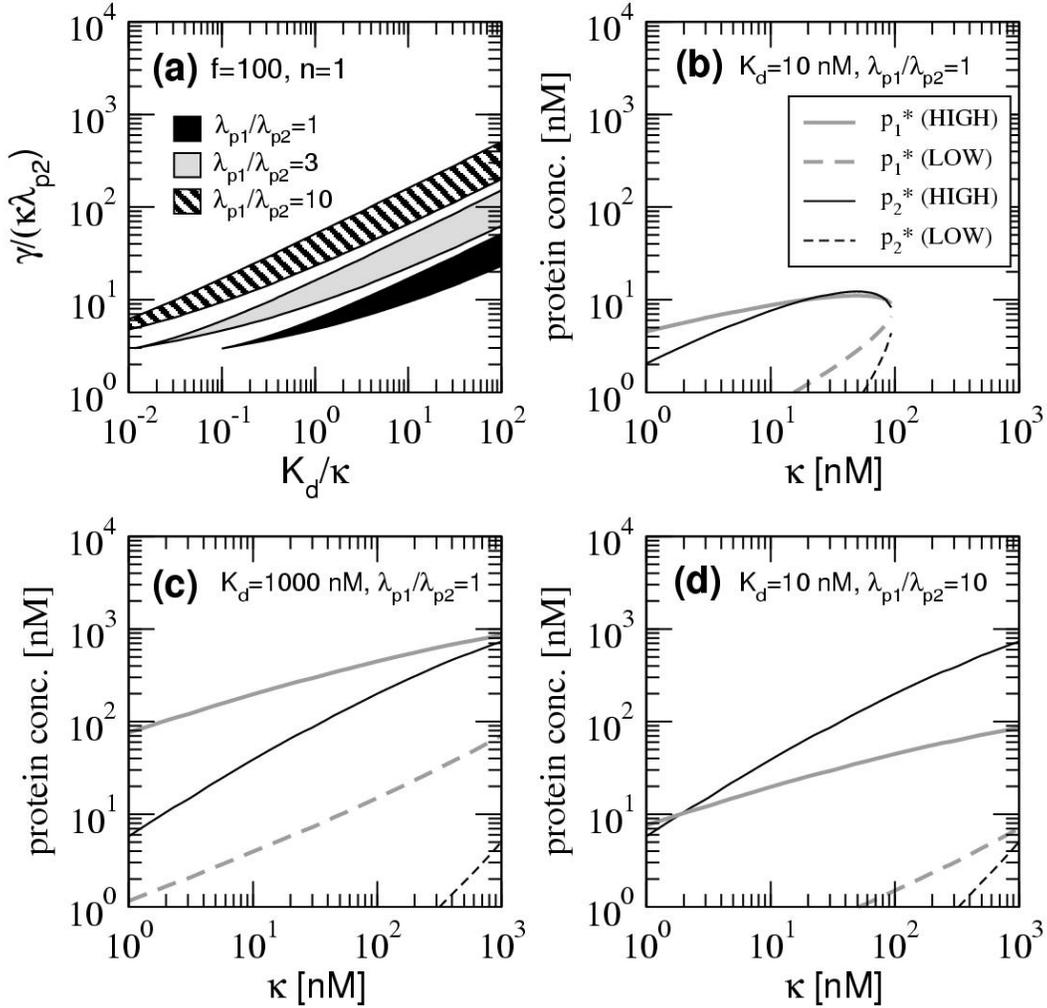

**Figure 4:** Quantitative characteristics of the bistable circuit with a single operator promoter ($n = 1$) and a strong activator ($f = 100$). (a) Regime of bistability in the parameter space for the circuit with linear degradation ($\lambda_{p1}/\lambda_{p2} = 1$) and with cooperative stability ($\lambda_{p1}/\lambda_{p2} > 1$). The axes show combinations of the parameters which are both useful for the discussion and natural in the quantitative description, see Supporting Information. (b) For linear degradation, the steady-state monomer (grey) and dimer (black) concentrations (i.e., $p_1^*$ and $p_2^*$) are plotted for different values of $\kappa$, with $K_d = 10$ nM and (c) $K_d = 1000$ nM, where $\gamma$ is chosen such that the system is in the middle of the bistable regime, i.e. the black band in (a), for each choice of $K_d$ and $\kappa$. For both $p_1^*$ and $p_2^*$, the solid curve is the concentration in the HIGH state and the dashed curve is the concentration in the LOW state. (d) Same plot as (b) for the circuit with cooperative stability ($\lambda_{p1}/\lambda_{p2} = 10$).

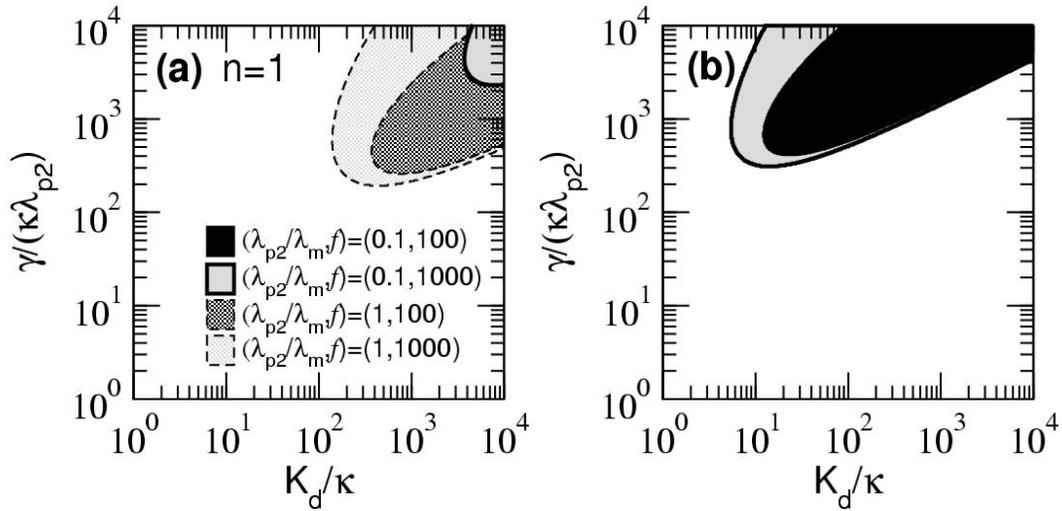

**Figure 5**: Quantitative characteristics of the three-gene oscillator involving repressor binding to a single operator site (*n=1*). (a) Oscillatory regime (shaded regions) in the parameter space for the case of linear degradation ($\lambda_{p1} = \lambda_{p2}$), using $\lambda_{p2}/\lambda_m = \{0.1, 1\}$ and $f = \{100, 1000\}$, representing {typical, rare} values, respectively (with the same parameter combinations on the axes as in Fig. 4). For $(\lambda_{p2}/\lambda_m, f) = (0.1, 100)$, sustained oscillation is not possible within the physiological parameter range being shown. (b) For the circuit with cooperative stability ($\lambda_{p1} = 10\lambda_{p2}$), we plot in the same parameter space the oscillatory regime with typical dimer/mRNA degradation rates ($\lambda_{p2}/\lambda_m = 0.1$), at the fixed repression strengths $f = 100$ (black) and $f = 1000$ (grey). A significant part of the accessible parameter space now displays oscillatory behavior.

# SUPPORTING INFORMATION

## *Dimer dissociation constant* in vitro *and* in vivo

The dimer dissociation constant $K_d$ is defined as the ratio of the concentrations of the monomers ($p_1$) and dimers ($p_2$) in steady state, i.e., $K_d = p_1^2/p_2$. In terms of the basic kinetic parameters, the association rate $k_a$ and the dissociation rate $k_d$, the *in vitro* dissociation constant is simply $K_d^{(0)} = k_d/k_a$. However, the effective value of $K_d$ is modified *in vivo* due to a variety of processes, including those that change the rate constants $k$. Even if the rate constants do not change, $K_d$ will also be affected by proteolysis or dilution, which alter the steady-state cellular protein concentrations. In the latter case, the combined effects of all these processes can be described at the level of chemical kinetics by the equations

$$\frac{dp_1}{dt} = \gamma - \lambda_{p1} \cdot p_1 - 2k_a p_1^2 + 2k_d p_2 \qquad [\text{S1}]$$

$$\frac{dp_2}{dt} = -\lambda_{p2} \cdot p_2 + k_a p_1^2 - k_d p_2 \qquad [\text{S2}]$$

where $\gamma$ is the protein synthesis rate and $\lambda_{p1}, \lambda_{p2}$ are the monomer, dimer degradation rates respectively; see the main text. The steady-state solution of Eq. S2 is $(\lambda_{p2} + k_d) \cdot p_2 = k_a p_1^2$, which according to the definition of the dissociation constant given above, yields an expression for the *in vivo* dissociation constant

$$K_d = \frac{k_d + \lambda_{p2}}{k_a} = K_d^{(0)} + \frac{\lambda_{p2}}{k_a}. \qquad [\text{S3}]$$

Note that the *in vivo* $K_d$ is independent of $\lambda_{p1}$, hence it does not depend on whether the protein is cooperatively stable (i.e., whether $\lambda_{p1}$ and $\lambda_{p2}$ are equal or different). As shown in Eq.

S3, $K_d$ is larger than the corresponding *in vitro* value $K_d^{(0)}$. Assuming that many dimers are diluted by cell division in the rapid exponential growth phase (~ 50 min half-life), then we can use typical monomer-monomer association rate of $k_a^{-1} \sim 20$ nM-min for small proteins (1, 2) to estimate a typical order of magnitude for the offset $\lambda_{p2}/k_a$ to be ~ 0.3 nM. Shifts of this magnitude in $K_d$ are marginally relevant for typical TFs with $K_d^{(0)} \sim 10$ nM, but they can become quite significant for larger proteins (e.g., enzymes) that have smaller $k_a$ or proteins that are rapidly degraded by proteolysis, see Table 1.

## *Thermodynamic models of transcriptional control*

In the text, we used the effective Hill functions

$$g_A([TF]) \approx \frac{f^{-1} + ([TF]/\kappa)^n}{1 + ([TF]/\kappa)^n} \qquad [S4]$$

$$g_R([TF]) \approx \frac{1 + ([TF]/\kappa)^n/f}{1 + ([TF]/\kappa)^n} \qquad [S5]$$

to describe the activities of the promoters shown in Fig. 1c-f. In this description, promoter activity is effectively characterized by three numbers, see Fig. 3: (i) $f$ which defines the maximum fold-change in promoter activity, (ii) $\kappa$ which indicates the concentration that separates the HIGH plateau from the transition region, and (iii) the "Hill coefficient" $n$ which describes the transcriptional cooperativity in the transition region. Here, we use a more realistic description of the promoter activity based on the thermodynamic models of transcriptional initiation (3,4,5), and relate the effective parameters used in Eqs. S4 and S5 to the biochemical parameters controlling protein-DNA and protein-protein interactions. We stress that our results which follow all implicitly presume that the promoters of interest are sufficiently weak (4,5).

**Simple activation and repression.** In the thermodynamic description, the promoter activity is assumed to be proportional to the equilibrium probability *P* that the RNA polymerase (RNAP) binds to the core promoter. The dependence of *P* on cellular TF concentrations (*TF*) for the four cis-regulatory architectures shown in Fig. 1 has been calculated elsewhere (4,5). The single activator (Fig. 1c) and repressor (Fig. 1e) promoters have the forms

$$P_{act1}([TF]) \propto \frac{1 + \omega_{A-P} \cdot [TF]/K_A}{1 + [TF]/K_A} \quad [S6]$$

$$P_{rep1}([TF]) \propto \frac{1}{1 + [TF]/K_R} + L \quad [S7]$$

In the above equations, $K_A$ and $K_R$ refer to the dissociation constant between the TF and the respective operator sequence *in vivo*, $\omega_{A-P} = \exp(-\Delta G_{A-P}/RT)$ is the Boltzmann weight of the activator-RNAP interaction, and *L* describes the effect of "promoter leakage" in repression which is expected to occur even for $[TF] \to \infty$ ‡.

Since the effective promoter activity functions ($g_A, g_R$) shown in Eqs. S4 and S5 are simply $P_{act1}$, $P_{rep1}$ normalized by their maximum values, one can compare Eqs. S4, S5 with S6, S7 to yield *n=1* for both cases, with $f = \omega_{A-P}$ and $\kappa = K_A$ for the activator, and $f = L^{-1}$ and $\kappa = K_R$ for the repressor. An important feature of the promoter activity function in the context of genetic circuits is the sensitivity *s*, which is defined as the absolute value of the log-log slope of $g([TF])$, i.e.,

$$s_A = \frac{d \ln g_A}{d \ln [TF]}. \quad [S8]$$

$$s_R = -\frac{d \ln g_R}{d \ln [TF]}. \quad [S9]$$

---

‡ For instance, this may be due to the collision of the replication fork with the repressed promoter which occurs at least once per cell cycle.

They are plotted in Fig. S1a as a function of $[TF]/\kappa$ for different values of $f=10, 100, 1000$ for the activator and the repressor. Note that the maximum value of $s([TF])$, referred to here as $s^*$, occurs at the mid point of the transition region (see Fig. 3) and is strongly dependent on the magnitude of the fold-change $f$. As shown in Fig. S1b, even at $f = 100$, s*=0.82 is still below the theoretical maximum set by $n = 1$. Since large $f$ is more readily available for repressors (due to the strong exclusion interaction between RNAP and the promoter-bound repressor), it is easier to attain large sensitivity using repressor-controlled promoters.

**Cooperative activation**. For the promoter shown in Fig. 1d involving two operators for the activator, the corresponding promoter occupancy probability (4,5) is:

$$P_{act2}([TF]) \propto \frac{1+[TF]/K_H + \omega_{A-P} \cdot [TF]/K_A + \omega_{A-A} \cdot \omega_{A-P} \cdot [TF]^2/(K_A \cdot K_H)}{1+[TF]/K_A + [TF]/K_H + \omega_{A-A} \cdot [TF]^2/(K_A \cdot K_H)} \qquad [S10]$$

where $K_H$ refers to the *in vivo* dissociation constant of the TF to upstream "helper" operator, and $\omega_{A-A}$ is the Boltzmann weight associated with activator-activator interaction. The form (S10) is clearly different from the Hill form (S4), and the values of the effective Hill parameters $n$ and $\kappa$ will necessarily depend on the actual values of the parameters $K_A$, $K_H$, $\omega_{A-P}$, and $\omega_{A-A}$. In the main text, this promoter was used to model the $P_{RM}$ promoter of phage lambda studied experimentally by Issacs et al (6). For this case, the primary and helper activator sites are the operators $O_{R2}$ and $O_{R1}$ respectively whereas we neglected the very weak repressive site $O_{R3}$ (7). The TF, lambda repressor protein CI, can stimulate transcription while bound at $O_{R2}$[§]; the same TF can interact cooperatively with an adjacent TF bound upstream at $O_{R1}$ (10).

---

[§] The stimulation of transcription by TFs (i.e. activation) typically described in thermodynamic models involves the "thermodynamic recruitment" of RNAP by the activator via a mutual attractive interaction (8). Transcriptional stimulation by CI involves instead catalyzing the rate of RNAP-promoter isomerization from a closed inactive form to an open active form (9). This change in isomerization rate, $\Delta k_{iso}$, is mathematically equivalent to "thermodynamic recruitment" with $\omega_{A-P} \approx \Delta k_{iso}$.

It is well known that the transcriptional stimulation provided by CI at P$_{RM}$ is weak, with $\omega_{A-P} \approx 11$ (9). The *in vivo* values of the CI-operator dissociation constants $K_A$ and $K_H$ are not known, but their ratio is expected to be equivalent to the ratio of the *in vitro* dissociation constants. The latter have been measured, together with the CI-CI affinity $\omega_{A-A}$. There are two sets of parameter values commonly used in the phage lambda literature: (i) Shea & Ackers (1985) (3) where $K_A/K_H \approx 17$ and $\omega_{A-A} \approx 25$, and (ii) Koblan & Ackers (1992) (11) where $K_A/K_H \approx 25$ and $\omega_{A-A} \approx 100$. In Fig. S2a, we plot the function $P_{act2}$ (normalized by its maximum value) against $[TF]/K_H$ for the two sets of parameters (the solid black and grey lines). The two curves are then fit to the effective Hill form (S4) to extract the effective parameters: $n \approx 1.53$ and $\kappa \approx 1.22 K_H$ for SA85; and $n \approx 1.69$ and $\kappa \approx 0.64 K_H$ for KA92; see the dashed lines in Fig. S2a. In Fig. S2b, we plot the sensitivity function $s_A([TF])$ for the two cases. We see that despite the improved Hill cooperativity for this promoter ($n \approx 1.5 - 1.7$), the marginal fold-change $f = 11$ in activation limits the maximum sensitivity to $s_A^* < 1$.

**Dual repression**. For the promoters controlled by two repressor sites (Fig. 1f), the corresponding activity function (4,5) is given by

$$P_{rep2}([TF]) \propto \frac{1}{(1+[TF]/K_{R1}) \cdot (1+[TF]/K_{R2})} + L \qquad [S11]$$

where $K_{R1}$ and $K_{R2}$ refer to the *in vivo* dissociation constants for the two operators, and $L$ again describes promoter leakage. Examples of such cis-regulatory constructs are the very strongly repressively controlled promoters constructed by Lutz & Bujard (12) with very low leakiness. In Fig. S3a, we plot the relative activity function for $K_{R1} = K_{R2}$, and $L = 10^{-3}$ (solid line). The dashed line is the best fit to the Hill form (S5), with the effective parameters $n = 1.63, \kappa \approx 0.52 K_{R1}$, and $f = L^{-1}$. The sensitivity function $s_R([TF])$ plays an important role in the ring-oscillator circuit (as will be shown below); it is plotted in Fig. S3b. Owing to the large $f$, $s_R^*$ is in this case close to its maximal theoretical value set by the Hill coefficient $n = 1.63$.

## *Bistability*

This genetic circuit shown in Fig. 1a consists of a single gene, which encodes a protein that homodimerizes and activates its own transcription. The steady-state occurs when degradation and synthesis rates are balanced, and is given by Eq. 4 or more conveniently by

$$\left(\frac{\gamma}{\kappa \cdot \lambda_{p2}}\right) \cdot g_A(p_2^*) = \frac{\lambda_{p1}}{\lambda_{p2}} \cdot \sqrt{\frac{K_d}{\kappa} \cdot \frac{p_2^*}{\kappa}} + 2\frac{p_2^*}{\kappa} \qquad [S12]$$

where $p_2^*$ is the steady-state TF dimer concentration. Given the form of the promoter activity function $g_A$, the above equation can be solved to yield the stable solutions $p_2^*/\kappa$ as a function of the dimensionless, programmable parameters, $\gamma/\kappa \cdot \lambda_{p2}$ and $K_d/\kappa$ for different choices of $\lambda_{p1}/\lambda_{p2}$. For the promoter containing a single operator site (Fig. 1c), $g_A(p_2^*)$ is given by Eqn. S4 with $n = 1$ as discussed above. Solving Eqn. S12 using Mathematica 4.2 (13), we obtain various regimes of parameter space supporting bistability. Such regimes are plotted in Fig. 4a for $f = 100$ and $\lambda_{p1}/\lambda_{p2} = 1, 3, 10$.

The above procedure is repeated for the promoter with *double* operators (Fig. 1d) which we use as a model of the $P_{RM}$ promoter studied in the experiment of Issacs et al (6). The corresponding promoter activity function is $g_A$ as sketched in Fig. S2a and described by the approximate Hill form (S4) with $f \approx 11$ and $n \approx 1.7$. In Fig. S4, we again plot the bistable region in the space of the two programmable parameters ($\gamma/\kappa \cdot \lambda_{p2}$, $K_d/\kappa$) for $\lambda_{p1}/\lambda_{p2} = 1$. Despite the improved Hill coefficient for this promoter when compared to the single operator promoter discussed above, the regime of bistability for both cases are similarly limited (compare the black band in Fig. S4 and Fig. 4a). This is a consequence of the small fold-change (*f*) involved in auto-activation of the $P_{RM}$ promoter studied by Isaacs et al. (6).

## *Oscillation*

For the repressillator circuit of Fig. 1b, we have 3 genes whose mRNA and (total) protein concentrations are denoted by $m^{(i)}$ and $p^{(i)}$ with $i \in \{1,2,3\}$. An approximate picture of the circuit behavior can be obtained by assuming that the 3 genes have the same properties, e.g., the same promoter structure and the same synthesis/turnover rates (14). The circuit topology of Fig. 1b then leads to the following kinetic equations

$$\frac{d}{dt} m^{(i)} = \alpha\, g_R\left(p_2^{(i-1)}\right) - \lambda_m m^{(i)} \qquad [S13]$$

$$\frac{d}{dt} p^{(i)} = \nu\, m^{(i)} - \Delta\left(p_2^{(i)}\right) \qquad [S14]$$

In Eq. S13, the promoter activity function $g_R\left(p_2^{(i-1)}\right)$ is given by Eq. S5 and the functional TFs are again dimers (of concentration $p_2^{(i)}$ for each species *i*), with $p_2^{(0)} = p_2^{(3)}$ completing the circuit loop. We assume rapid equilibration between the monomers and dimers, such that the dimer and total protein concentrations are related by the condition

$$p = \sqrt{K_d \cdot p_2} + 2p_2 \qquad [S15]$$

for each species. According to the spirit of our approximation, these dimers all have the same $K_d$. In Eq. S14, we introduced the protein degradation function

$$\Delta = \lambda_{p1}\sqrt{K_d \cdot p_2} + 2\lambda_{p2} p_2 \qquad [S16]$$

which gives the total protein degradation rate.

To find the condition for oscillation for the system defined by Eqs. S13-S16, we follow the analysis of Elowitz (15) and first solve for the steady-state concentrations $\{m^{(i)} = m^*, p_2^{(i)} = p_2^*\}$ such that the left-hand side of Eqs. S13-S14 are zero. This is given by the condition

$$\gamma \cdot g_R(p_2^*/\kappa) = \Delta(p_2^*) \quad [S17]$$

where $\gamma = \dfrac{\alpha \cdot v}{\lambda_m}$. We then analyze small perturbations about this steady state and find the condition where undamped oscillatory solutions emerge (16). This amounts to finding the purely imaginary eigenvalues of the following Jacobian**:

$$J = \begin{pmatrix} -1 & 0 & 0 & 0 & 0 & -x \\ 1 & -y & 0 & 0 & 0 & 0 \\ 0 & -x & -1 & 0 & 0 & 0 \\ 0 & 0 & 1 & -y & 0 & 0 \\ 0 & 0 & 0 & -x & -1 & 0 \\ 0 & 0 & 0 & 0 & 1 & -y \end{pmatrix} \quad [S18]$$

where $x = -\dfrac{\gamma}{\lambda_m} \cdot \dfrac{d}{dp} g_R(p_2(p))\bigg|_{p^*}$, $y = \lambda_m^{-1} \cdot \dfrac{d}{dp} \Delta(p_2(p))\bigg|_{p^*}$, and $p_2(p)$ is given by Eq. S15.

The only acceptable solution of this system is given by the condition

$$\dfrac{(y+1)^2}{y} = \dfrac{3(x/y)^2}{4 - 2(x/y)} \quad [S19]$$

which can be alternatively expressed as $x/y = \psi(y)$ where

$$\psi(y) = \dfrac{(1+y)^2}{3y}\left(\sqrt{1 + \dfrac{12y}{(1+y)^2}} - 1\right). \quad [S20]$$

It is straightforward to show that the steady-state solution is *stable* whenever $x/y < \psi(y)$. Thus oscillation is not possible in this regime. For $x/y > \psi(y)$, we verified that the system remains in an oscillatory state by direct numerical integration of the kinetic Eqs. S13 and S14.

---

** We have rewritten Eqs. S13-S14 where the mRNA concentration and time have been rescaled to $m \cdot v/\lambda_m$ and $t \cdot \lambda_m$.

The condition $x/y = \psi(y)$ defines the boundary of the oscillatory phase. The function $\psi(y)$ is plotted in Fig. S5 and the oscillatory region is shaded.

From the form of $\psi(y)$, it is clear that no oscillation is possible if $x/y < 4/3$. For $x/y > 2$, oscillation will occur independently of $y$. For the intermediate range $4/3 < x/y < 2$, oscillation is most favorable for $y \approx 1$, i.e. the minimum of $\psi(y)$. To interpret the physical meaning of these conditions, it is useful to note the relation[††]

$$\frac{x}{y} = \frac{s_R(p_2^*)}{r(p_2^*)} \qquad [S21]$$

where $s_R \equiv -\left.\frac{d \ln g_R}{d \ln p_2}\right|_{p_2^*}$ is the sensitivity function, and $r \equiv \left.\frac{d \ln \Delta}{d \ln p_2}\right|_{p_2^*}$ is the log-log slope of the degradation function defined in Eq. S16. Eq. S21 shows that the function (i.e. oscillation) of the circuit is intimately related to the sensitivity. Since $1/2 \leq r \leq 1$, we conclude that oscillation will always occur if $s_R(p_2^*) > 2$. However, this is hardly satisfied for typical promoters. On the other hand, we note that in the regime where $r \approx 1/2$ (i.e., the degradation flux is predominantly through monomer loss), oscillation will always occur if $s_R(p_2^*) > 1$. This sensitivity is achievable by a promoter with two repressor sites (see Fig. S3b), but not for a promoter with a single repressor site since $s_R < n = 1$ (Fig. S1). For the latter promoter, oscillation can still occur if $s_R(p_2^*) > 2/3$ and $y$ is close to 1 (see Fig. S4).

The value of $y$ is given explicitly by the model parameters as

$$y = \left(\frac{\lambda_{p1}}{\lambda_m}\sqrt{\frac{K_d}{p_2^*}} + 2\frac{\lambda_{p2}}{\lambda_m}\right) \bigg/ \left(\sqrt{\frac{K_d}{p_2^*}} + 2\right) \qquad [S22]$$

---

[††] This relation can be derived starting from the definition of $x$ and $y$, by applying the chain-rule of differentiation and invoking the steady-state condition (S17).

There are three regimes for $y$: (i) when $p_2 = K_d$ (i.e. where monomers are the dominant species), then $y = \lambda_{p1}/\lambda_m$, and oscillation is most favorable in this regime if the mRNA and *monomer* degradation rates are comparable (where $y \approx 1$). (ii) when $K_d = p_2^* = (\lambda_{p1}/\lambda_{p2})^2 K_d$ (i.e. dimers are the dominant species, but degradative flux is still predominantly through the monomers), then $y \approx \frac{\lambda_{p1}}{\lambda_m}\sqrt{K_d/p_2^*}$, and depending on the steady-state dimer concentration $p_2^*$, $y$ spans the range $\lambda_{p2}/\lambda_m < y < \lambda_{p1}/\lambda_m$. (iii) when $p_2^* \gg \left(\frac{\lambda_{p1}}{\lambda_{p2}}\right)^2 K_d$ (dimers are the dominant species and degradative flux is predominantly through dimers), then $y = \lambda_{p2}/\lambda_m$. From the above analysis, we see that cooperative stability (i.e., $\lambda_{p1}/\lambda_{p2} > 1$) improves the condition for oscillation by increasing the regime where degradation flux is monomer dominated, so that a large regime of parameter space is governed by $r \approx 1/2$ and $y \approx 1$.

This is illustrated explicitly in Fig. 5, where we plot the region of spontaneous oscillation as a function of the programmable parameters ($\gamma/\kappa \cdot \lambda_{p2}$, $K_d/\kappa$) for different fixed ($\lambda_{p1}/\lambda_{p2}$, $\lambda_{p2}/\lambda_m$) using $g_R$ for the promoter containing a single repressor site with $n = 1$ and $f = 100, 1000$. The contour of oscillatory parameter space shown in Fig. 5 was generated using Mathematica 4.2 (13) by first numerically solving Eq. S17 to obtain $p_2^*$ for each set of parameters, then determining $y$ and $x/y$ according to their definitions, e.g., see Eq. S18, and finally comparing to the oscillatory condition $x/y > \psi(y)$.

44

| Protein | $K_d^{(0)}$ (nM) | $k_d^{-1}$ (min) | $k_a^{-1}$ (nM-min) | $K_d$ (nM) | References |
|---|---|---|---|---|---|
| Arc | 10 | 0.2 | 2 | 10 | (17) |
| HIV-1 protease | 4 | 8 | 32 | 4.5 | (18) |
| CRP | 0.1-1 | 330 | 30 - 300 | 0.5 – 5 | (19,20) |
| β-galactosidase | 0 | 0 | 3900 | 56 | (21) |

**Table 1**: *In vitro* values for $K_d^{(0)}$, $k_d^{-1}$, $k_a^{-1}$ were taken from the literature for a few exemplary cases. Presuming that dimer "turnover" *in vivo* occurs primarily through dilution $\lambda_{p2}^{-1} \approx 70$ min, we can estimate the *in vivo* dimer dissociation constant $K_d$. Note that two-state dimers such as the Arc repressor tend to have smaller $k_a^{-1}$ than the typical value given above ($k_a^{-1} \sim 20$ nM-min), due presumably to the lack of orientation constraints in the association process (22). On the other hand, large proteins such as β-galactosidase will tend to have larger $k_a^{-1}$, which can lead to large shifts in $K_d$.

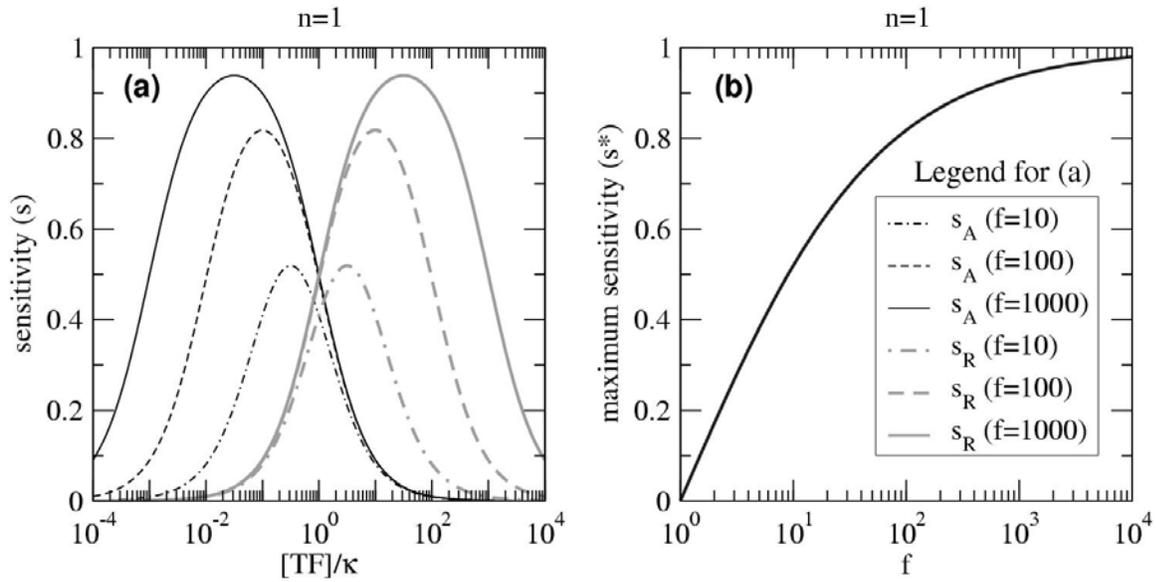

**Figure S1:** (a) Plot of the sensitivity for simple activation ($s_A$) and repression ($s_R$) at n=1 for different $f = 10, 100, 1000$. The legend to these curves is located in the adjacent figure. (b) Plot of the maximum sensitivity ($s^*$) as a function of promoter strength $f$. The shape of the curve is independent of $n$ (i.e. the Hill coefficient simply scales the height or $s^*$) and whether the Hill function is repressive ($g_R$) or activating ($g_A$).

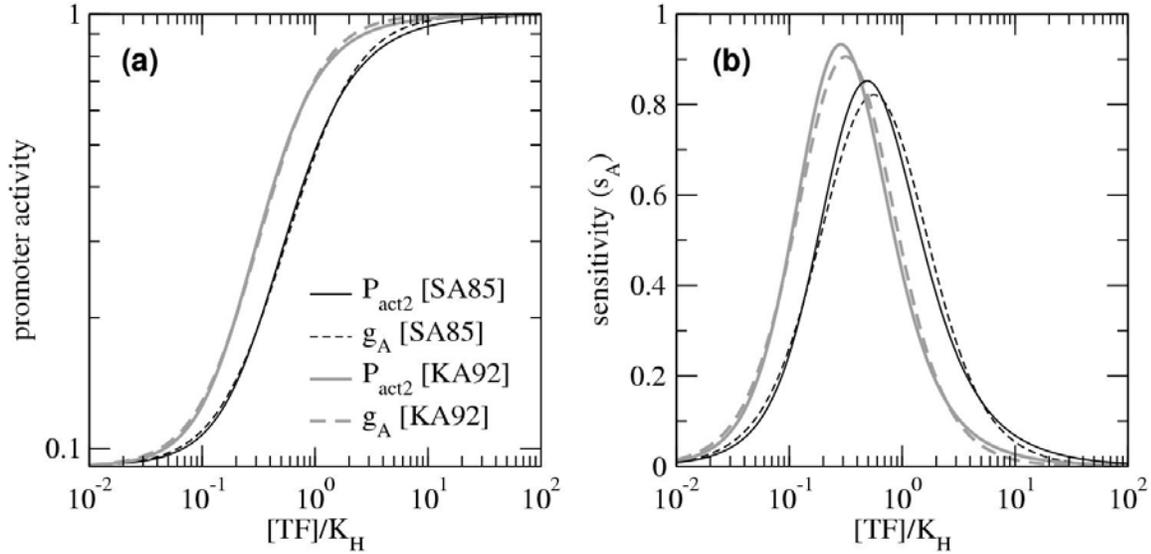

**Figure S2:** (A) Log-log plot of the promoter activity $P_{act2}$ for the Isaacs promoter (6) (see Fig. 1d) as a function of relative concentration $[TF]/K_H$ using two different sets of experimental parameters: SA85 (3) and KA92 (11). Each set of parameters is fit to $g_A([TF])$, an approximate Hill function described in Eq. S4. We obtain: $n \approx 1.53$ and $\kappa \approx 1.22 K_H$ for SA85; and $n \approx 1.69$ and $\kappa \approx 0.64 K_H$ for KA92 (B) The log-log slope of the activity function (the sensitivity $s_A$) as a function of relative protein concentration $[TF]/K_H$ is plotted for both sets of parameters for both $P_{act2}$ and approximate Hill function $g_A([TF])$.

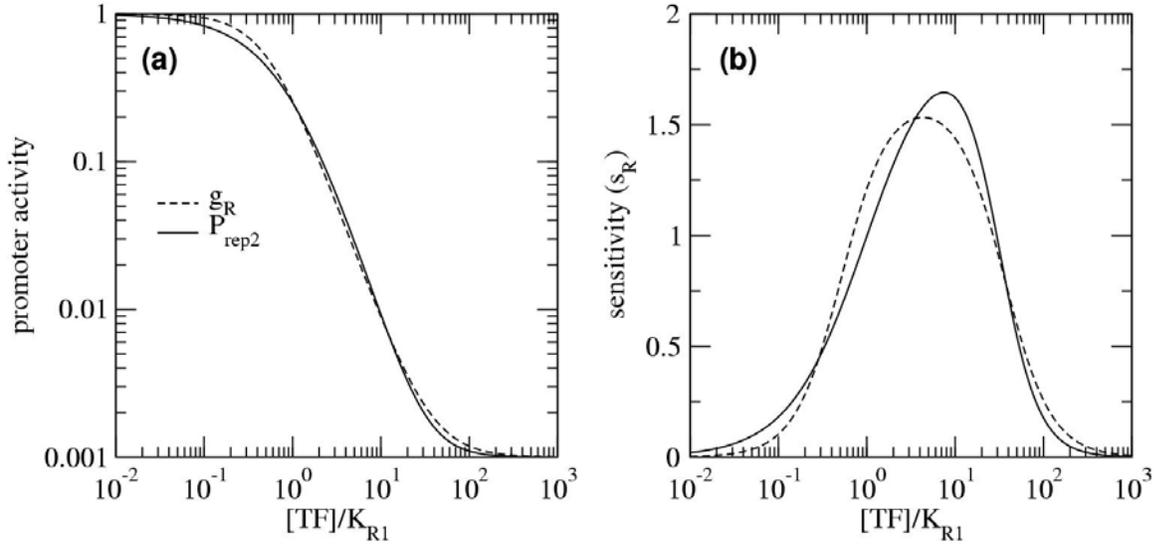

**Figure S3:** (A) Log-log plot of the promoter activity $P_{rep2}$ that describes the promoters engineered by Lutz et al. (12) (see Fig. 1f) as a function of relative concentration $[TF]/K_{R1}$ using the experimental parameters $K_{R1} = K_{R2}$ and $L = 10^{-3}$. The promoter activity $P_{rep2}$ is fit to $g_R([TF])$, an approximate Hill function described in Eq. S5. We obtain parameters $n = 1.63, \kappa \approx 0.52 K_{R1}$. (B) The log-log slope (sensitivity $s_R$) of the promoter activity as a function of relative protein concentration $[TF]/K_{R1}$ is plotted for $P_{rep2}$ and approximate Hill function $g_R([TF])$.

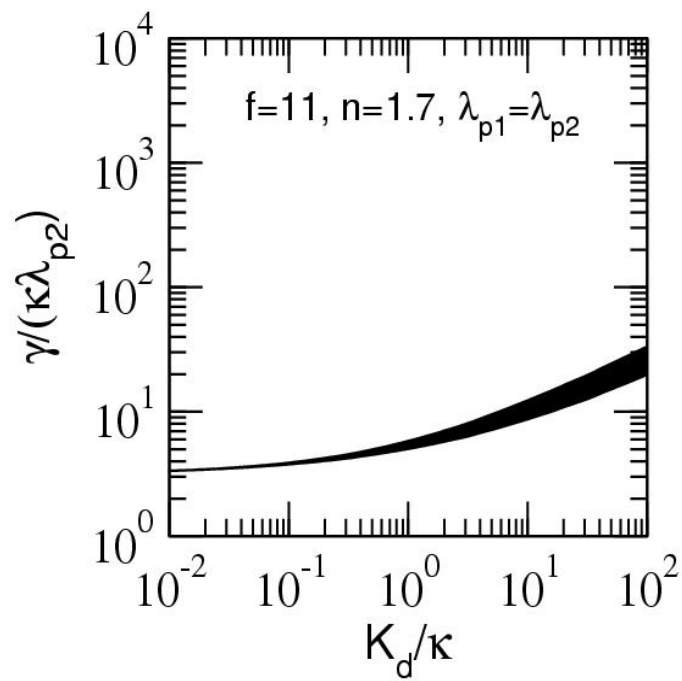

**Figure S4:** The regime of bistability (shaded region) in the parameter space ($\gamma/\kappa \cdot \lambda_{p2}$, $K_d/\kappa$) for linear degradation ($\lambda_{p1} = \lambda_{p2}$) using the parameters $f = 11$, $n = 1.7$ (see Fig. S2) that best describe the $P_{RM}$ promoter used by Isaacs et al. (6).

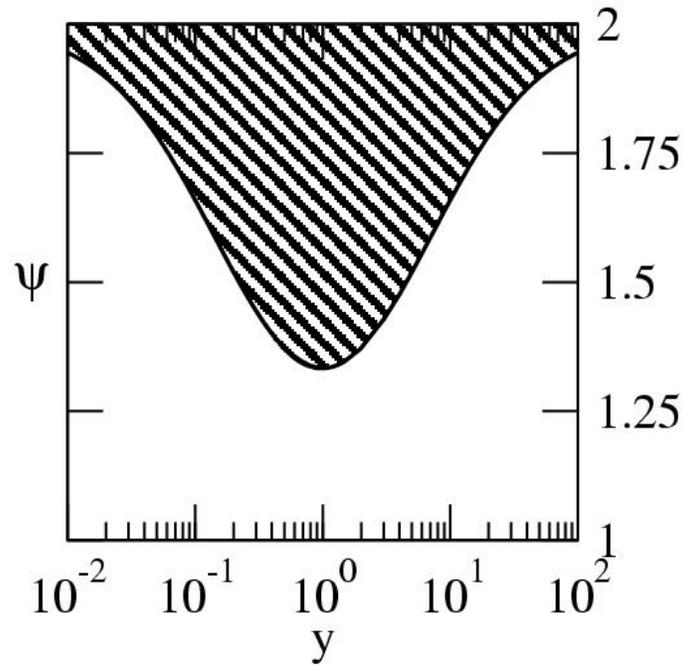

**Figure S5:** Plot of $\psi(y)$ where the unstable solutions (oscillatory) that satisfy $x/y > \psi(y)$ are in the hatched region. All $x/y > 2$ are oscillatory, independent of the relative rates of protein and mRNA degradation encapsulated by the parameter $y$.